\documentclass[sigconf]{acmart}

\let\savedbaselinestretch\baselinestretch
\usepackage{listings}
\usepackage{color}
\usepackage{makecell}
\usepackage[shortlabels]{enumitem}
\definecolor{dkgreen}{rgb}{0,0.6,0}
\definecolor{gray}{rgb}{0.5,0.5,0.5}
\definecolor{mauve}{rgb}{0.58,0,0.82}

\graphicspath {{img}}
\let\baselinestretch\savedbaselinestretch
\AtBeginDocument{%
  }

\begin{document}

\title{Microservices Are Dying, A New Method for Module Division Based on Universal Interfaces}

\author{Qing Wang}
\affiliation{%
  \institution{BNRist, Tsinghua University}
  \city{Beijing}
  \country{China}}
\orcid{0009-0003-0446-577X}    
\email{wangqing.2005@tsinghua.org.cn}  

\author{Yong Zhang}
\affiliation{%
  \institution{BNRist, Tsinghua University}
  \city{Beijing}
  \country{China}}
\orcid{0000-0001-8803-2055}  
\email{zhangyong05@tsinghua.edu.cn} 

\begin{abstract}
Although microservices have physically isolated modules, they have failed to prevent the propagation and diffusion of dependencies. To trace the root cause of the inter-module coupling, this paper, starting from the impact assessment approach for module changes, proposes a conceptual method for calculating module independence and utilizes this method to derive the necessary conditions for module independence. Then, a new system design philosophy and software engineering methodology is proposed, aimed at eliminating dependencies between modules. A specific pattern is employed to design a set of universal interfaces, serving as a universal boundary between modules. Subsequently, this method is used to implement a platform architecture named EIGHT, demonstrating that, as long as module independence is guaranteed, even a monolithic application within a single process can dynamically load, unload, or modify any part at runtime. Finally, the paper concludes that this architecture aims to explore a novel path for increasingly complex systems, beyond microservice and monolithic architectures.
\end{abstract}



\maketitle

\section{Introduction}
Since microservices were formally proposed in 2014\cite{microservice2014,Schwartz2017Microservices}, they have been rapidly practiced and popularized in major companies \cite{Wang2021PromisesAC}, because they can decompose large complex systems, effectively organize multi-team development, rapidly respond to changes \cite{SalaheddinELGHERIANI2022MICROSERVICESVM,Ksling2017AgileDO}. Today, more and more companies are migrating their businesses to microservice architectures\cite{Sylemez2022ChallengesAS,Buchgeher2017MicroservicesIA}. However, as microservices become increasingly widely used, some negative effects have started to appear. Increased system complexity  \cite{Soldani2018ThePA,Viggiato2018MicroservicesIP}, data consistency difficulties  \cite{Ajoux2015ChallengesTA,Velepucha2021MonolithsTM}, performance degradation \cite{Raharjo2022ReliabilityEO}, difficulty in troubleshooting and debugging \cite{Dragoni2016MicroservicesYT}, and high implementation and maintenance costs \cite{Gos2020TheCO}are problems faced by many practitioners. In fact, the migration to microservices may be a burden for small and medium-sized enterprises\cite{Baskarada2018ArchitectingMP}.

The main purpose of enterprises using microservices is to solve the problem of module division and conquer after the growth of system scale, but the module independence of microservices also has problems. For example, public service upgrades leading to failures (notably, a recent Twitter system-wide failure caused by a service interface modification \cite{twitter2023} and the massive service outage caused by a null pointer in Google's API management system \cite{google2025}), and the accumulation of historical service APIs \cite{Lercher2023MicroserviceAE}, all of which show that microservice modules are not independent. Another difficulty lies in the measurement of module independence. The current measurement of module coupling, whether based on static code analysis \cite{Sangal2005UsingDM,MacCormack2006ExploringTS} or real-time traffic collection \cite{Zheng2023ChainDetAT}, is flawed. For example, how to explain the obvious difference in modularity between microservices and monolithic architectures, which also use interface-based programming to split modules? 

This paper will first analyze the dependencies between modules, propose a conceptual metric for measuring module coupling degree and impact scope, and use this metric to derive conditions for module independence. Subsequently, based on the implicit meaning of module independence conditions, the root causes of module coupling are identified. An architecture is then proposed, elucidating its philosophical design principles and module division principles, and using the previous evaluation metrics to illustrate its superior module independence. Furthermore, based on this conceptual model, a highly dynamic monolithic component architecture named EIGHT is implemented, which adopts a set of universal interfaces as the boundary of module division. A platform and demonstration cases are provided to prove that, with a reasonable architecture and development approach, even a monolithic application can dynamically change its form and modify any part of the system at runtime, similar to microservices. This architecture offers advantages in terms of module independence, development agility, deployment flexibility, and maintenance convenience. It has lower environmental requirements, less resource consumption, and lower implementation costs, providing a new path for developers facing the dilemma of choosing between monolithic applications and microservices.

The paper will continue by introducing a conceptual modularity metric and deriving conditions for module independence in Section 2. Then, it will propose a conceptual architecture designed based on a new worldview in Section 3. Section 4 will introduce the platform system and application practices developed from this conceptual architecture. Following that, Section 5 will discuss related work. The paper will conclude with a summary in Section 6.

\section{Impact Scope: A Conceptual Metric for Module Changes}
To evaluate the impact of module changes on the application, this research defines a metric, which are named Impact Scope Metric (ISM), and elaborates on module independence in this way. This metric is a conceptual metric, and the main purpose is to construct a system to analyze the way of module coupling and related influencing factors, as well as to reveal the differences in the diffusion of changes under different architectures, and to study the characteristics of what kind of modules are truly independent, and ultimately to derive the basic characteristics of an ideal system.

\subsection{Definitions}
This method divides the software into three levels: application, module, and service, considering an application to be composed of a set of modules and a module to be composed of a set of services. The following definitions are provided for them:

\noindent\textbf{Definition 1}. $Application$: 

An application is an executable whole that consists of a set of modules. The symbolic representation of the application is $A_{aname}$, where $aname$ is the name of the application and is globally unique.

\noindent\textbf{Definition 2}. $Module$: 

A module is a whole that consists of a set of services. In modular architectures such as microservices or OSGi, a module is the smallest executable unit. Modules can be reused in different applications but are unique within the application. The symbolic representation of a module is $M_{mid}$, where $mid = aname.mname$. the $aname$ denotes the name of the application to which it belongs, and the $mname$ denotes the name of the module itself. The $mid$ is globally unique. 

\noindent\textbf{Definition 3}. $Service$: 

A part of a module that can impact other modules. It is the smallest unit of change. Different architectures have different types of services. For example, in object-oriented programming, public properties and methods accessible to other classes, in microservices, service interfaces of a module. The symbolic representation of a service is $S_{sid}$, where $sid = mid.sname$, $mid$ is the module id to which the service belongs, and $sname$ is the service name, which is unique within the current module. The $sid$ is globally unique. Each module has a special $sid = mid.self$, where $self$ refers to the module itself. Any change in a module, whether it alters the part that interacts with other modules (such as accessible member variables, methods, interfaces, services, etc.), $mid.self$ is included in the impact scope.

\begin{figure}[htbp]
  \centering
  \includegraphics[width=0.6\linewidth]{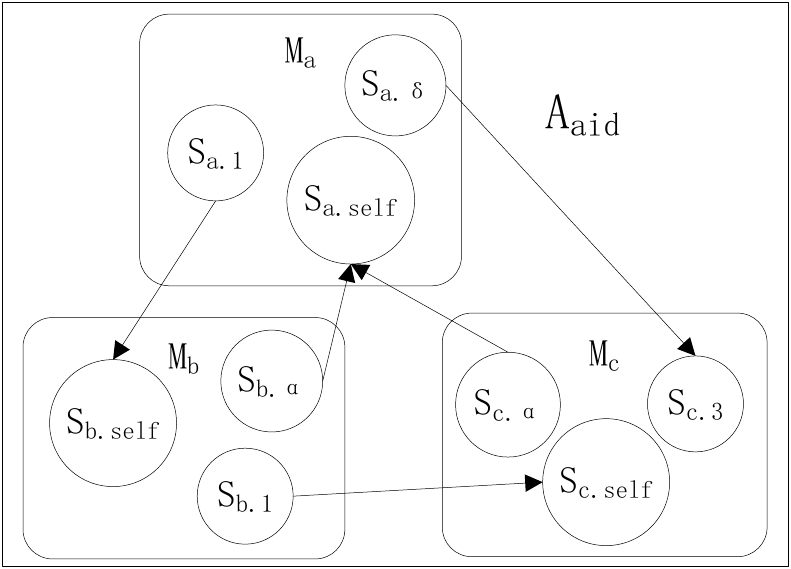}
  \caption{Application Structrue}
  \label{fig:application-structrue}  
\end{figure}

An application should be represented as a tree structure in terms of its hierarchical structure. However, in terms of mutual impact and association, it should be represented as a network structure. Numerous researches support this model\cite{ubelj2012SoftwareST,Myers2003SoftwareSA,Cai2009SoftwareEP,Chong2015AnalyzingMA}. These association relationships, according to Definition 3, are established through services. The coupling between modules and applications is reflected in the association and mutual impact of services. Therefore, the structure of a typical application is shown in Fig.~\ref{fig:application-structrue}.

The relationships between applications, modules and services can be analyzed in terms of set, and defines a set of functions to decompose the set of applications into the set of modules ($a_m(\{A\})$), decompose the set of modules into the set of services ($m_s(\{M\})$), decompose the set of applications into the set of services ($a_s(\{A\})$), or merge the set of services into the set of modules ($m_a(\{M\})$), or merge the set of modules into the set of applications ($m_a(\{M\})$) and merge the set of services into the set of applications ($s_a(\{S\})$). Due to the finite number of modules contained in any application and the services contained in a module, it is easy to prove that both the input and output of the above functions are finite sets.

\noindent\textbf{Definition 4}. $Changes$: 

A set of events that impact one or more services, causing their state to change. Any change to applications and modules ultimately manifests as changes to services. In this research, these changes are categorized into three types of impact scopes.
\begin{enumerate}[leftmargin=*,label=\textbf{Type \arabic*}.]
\item {\texttt{Static changes}}: Changes in the static content, structure, organization, etc. of the application, including code modifications, module interface adjustments, design refactoring, etc. The category is denoted as $s$.
\item {\texttt{Runtime changes}}: Changes caused by environmental and state changes during application runtime, such as node and service failures, interruptions in transactions or loss of data due to restarts caused by updates, etc. The category is denoted as $r$.
\item {\texttt{Non-runtime changes}}: Changes that are neither static nor runtime changes but can impact other modules, such as triggering events such as recompilation, packaging, deployment, etc., also include the correlation impacts brought by the development process, team structure, and project management approach. The category is denoted as $o$.
\end{enumerate}

\noindent\textbf{Definition 5}. $Rules$: 

A set of rules that cause associated services to change due to a specific service change, expressed as $R(x)$, where the independent variable $x$ is the set of contexts, usually one or more of the three types of changes described above, and the return value is the set of rules in the current context. A rule is defined as $(\exists S_i)(\exists S_j)(c(\{s_i\})\to c(\{s_j\}))$, then $c(\{s_i\})$ will cause $c(\{s_j\})$ to occur. 

Rules are context-dependent. Rule sets in different contexts are different. For example, for the same modification of a module, the module may be independent when examining static changes, but when examining non-runtime changes, changes in all other modules may cause it to change.

\noindent\textbf{Definition 6}. $Impact$: 

A set of changes caused by the current rules and changes, defined as the function $p(c(\{S\}),R(x))$. Its parameters $c(\{S\})$ are a set of changes, $R(x)$ is the set of rules in the determined context, and the function's return value is also a change set, whose elements are all changes caused by these changes in the current context.

Define $p_d(c(\{S\}),R(x))$ as the set of direct changes caused by a set of changes. Direct changes are those whose elements are both and only associated with $c(\{S\})$. Then $p(c(\{S\}),R(x)) = c(\{S\}) \cup p(p_d(c(\{S\}),R(x)),R(x))$, which can be obtained recursively using $p_d(c(\{S\}),R(x))$ to obtain $p(c(\{S\}),R(x))$.

\noindent\textbf{Definition 7}. \textit{Impact-Scope}: 

The function $scope(c(\{S\}))$ is called the impact-scope function, which takes a set of changes to services as input and outputs a set of services that have changed. The impact-scope of a change in a given context is $scope(p(c(\{S\}),R(x)))$. It follows that the modules impacted by a change are $s_m(scope(p(c(\{S\}),R(x))))$, and the  applications impacted by a change are $s_a(scope(p(c(\{S\}),R(x))))$.

Notably, this function can track the impact scope of a specific change $scope(p(c(\{S_{any}\}),R(x)))$, or other modules impacted by a module change $s_m(scope(p(c(\{M_{any}\}),R(x))))$.

\begin{figure}[htbp]
  \centering
  \includegraphics[width=0.9\linewidth]{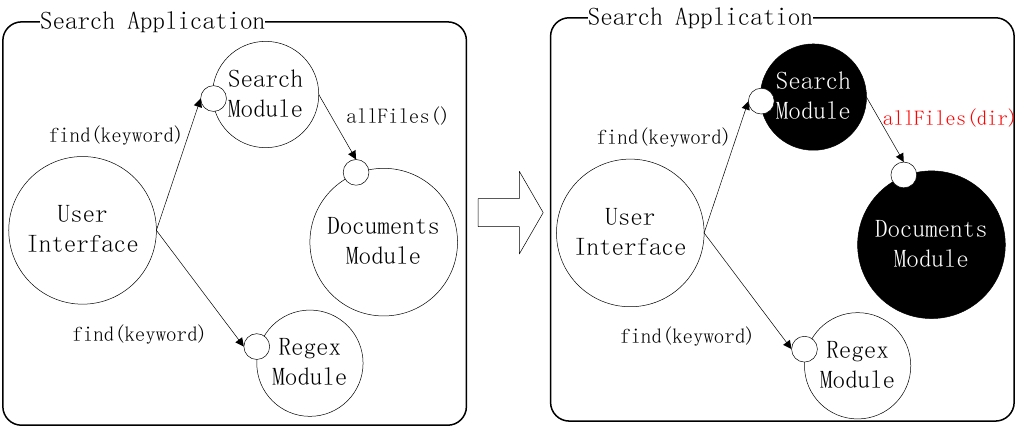}
  \caption{Application Refactoring}
  \label{fig:application-refactoring}  
\end{figure}

Using the above metrics for analysis, it can be seen that the primary advantage of microservices architecture over monolithic architecture lies in the differing impact scopes when changes occur in different contexts. For example, consider a search application with several modules: Document, Search, Regex, and UserInterface. The Document module's interface is called by the Search module. When the Document interface changes (from $allFiles()$ to $allFiles(dir)$), the Search module's code also needs to be modified, while the UserInterface and Regex modules do not require code changes. As shown in Fig.~\ref{fig:application-refactoring}.

This scenario is based on the above definition. can be expressed as for: $scope(p(c(\{M_{Document}\}),R(\{s\}))) = \{M_{Document}, M_{Search}\}$. That is, in a static change context, the impact scope of changes to $M_{Document}$ is $M_{Document}$ and $M_{Search}$. but for a monolithic application, where each new code commit causes all module developers to pull, compile, and test the code, the impact scope is significantly different in context $o$. In this case, the scope of the application with the microservices architecture remains the same, while the scope of the monolithic application is: $scope(p(c(\{M_{Document}\}),R(\{o\}))) = \{M_{Document}, M_{Search}, M_{Regex}, M_{UserInterface}\}$.

\subsection{The Significance of ISM}
The following will discusses what the previous metrics represent in software engineering and draws some important inferences. First, set a few definitions:

\noindent\textbf{Definition 8}. \textit{Independence}: \begin{equation}\begin{aligned}(\exists M_i)(\exists M_j)(s_m(scope(p(c(\{M_i\}), R(x)))) \\ \cap \{M_j\} = \oslash)\end{aligned}\end{equation}
then $M_i$ is considered independent of $M_j$ in the context set $x$. This intuitively means that the impact scope of $M_i$ in this change does not include $M_j$, so it will not impact $M_j$. 

\noindent\textbf{Definition 9}. \textit{Completely Independence}: \begin{equation}\begin{aligned}(\forall R(x))(\exists M_i)(\exists M_j)(s_m(scope(p( c(\{M_i\}),R(x)))) \cap \\ \{M_j\} = \oslash)\end{aligned}\end{equation}
then $M_i$ is considered completely independent of $M_j$. This intuitively means that any change that occurs to $M_i$ under any rule will not impact $M_j$.

\noindent\textbf{Definition 10}. \textit{Absolutely Independence}: \begin{equation}\begin{aligned}(\forall R(x))(\exists M_i)(\exists A_j)(M_i \in a_m(\{A_j\}) | s_m( scope(p(\\c(\{M_i\}), R(x)))) \cap a_m(\{A_j\}) = \{M_i\})\end{aligned}\end{equation}
then $M_i$ is considered absolutely independent of its current application. This intuitively means that any change to $M_i$ will not impact other modules in the application.

Therefore, the following theorem and corollaries can be derived:

\noindent\textbf{Theorem 1}. When $M_i$ is completely independent of $M_j$, $M_j$ does not need to know any information about $M_i$ to be implemented. 

\noindent\textbf{Proof}. Assume that $M_j$ must know some information about $M_i$ to be implemented. However, any change to $M_i$ will not affect the implementation of $M_j$. Therefore, changes to the information $M_j$ needs to know will not affect $M_j$. This means that whether $M_j$ knows this information or not will not affect the implementation, contradicting the assumption. Therefore, the proof is complete. $\square$

\noindent\textbf{Corollary 1}. When $M_i$ is completely independent of $M_j$, $M_j$ can be implemented before $M_i$.

\noindent\textbf{Proof}. Since $M_j$ does not need to know any information about $M_i$ to be implemented, the implementation of $M_j$ is not affected by the implementation of $M_i$, and it can be implemented before $M_i$. $\square$

\noindent\textbf{Definition 11}. \textit{Ideal System}: 

When $M_i$ and $M_j$ are completely independent of each other, $M_j$ and $M_i$ can be implemented in parallel. When $M_i$ is absolutely independent, $M_i$ can be implemented arbitrarily without impacting the current application. When all modules in the application are absolutely independent, any part of the application can be implemented independently without impacting the current application, which is called an ideal system. 

It can be easily demonstrated that any architecture based on interface-based programming (including monolithic architectures that directly call methods and interfaces, OSGi which utilizes in-process service interface calls, and microservices architectures that employ RPC), is not an ideal system. This is because the caller must, at a minimum, be aware of the interface information to proceed with development, and once the interface changes, the caller module must also make corresponding changes.

\noindent\textbf{Corollary 2}. The necessary condition for constructing an ideal system is that the developer of each module can be completely unaware of the implementation details of any other module.

\noindent\textbf{Proof}.  According to Theorem 1, mutual ignorance between modules is the necessary condition for complete independence between modules. It follows that the degree of understanding a developer has of other modules is positively correlated with the coupling between modules. Therefore, the necessary condition for constructing an ideal system is that is that the developer of each module can be completely unaware of the implementation details of any other module. This corollary is intuitive and self-evident. $\square$

\section{The Ideal System}

The analysis of those metrics and corollaries reveals that the coupling between modules is directly related to the module developers' knowledge of other associated modules; the more information is known, the deeper the coupling. The definition of the ideal system, however, requires that developers perform development without knowledge of any information beyond the functionality of their own module. To achieve this seemingly impossible mission, distinct from interface-based programming, this research has designed a new architecture-\textbf{modules are connected to each other and interact through connections on which arbitrary functions can be run}-and has named it substance-connection model.

\subsection{Substance-Connection Model (SCM)}

Before explaining how modules can be developed independently without knowledge of one another, and why the aforementioned architecture can realize the ideal system, it is necessary to briefly outline the problems within current programming languages and system design philosophies, and why they lead to widespread dependencies and difficulty adapting to change.

As the scale of software continues to expand, how to effectively organize larger development teams to develop more complex software has always been one of the core problems in the field of software engineering. Therefore, programming languages have evolved toward modeling the real world, aiming to divide complex systems into distinct modules based on human perception of reality. Object-Oriented Programming (OOP) is a design philosophy widely adopted in recent decades for modeling the real world. It uses inheritance to express the relationships between various entities in the real world and achieves polymorphism through interfaces\cite{Booch1986ObjectorientedD}. This is similar to the fundamental structure Spinoza established for the world in $Ethics$\cite{deSpinoza2000DESE}, using substance, attributes, and modes. However, OOP's approach to modeling the real world is inherently problematic. Minor changes in software often result in an uncontrollable impact scope. In contrast, the real world adjusts and adapts to these changes at runtime, restricting them to a very small scope. Where exactly does the difference lie?

Imagine a simple scenario: one day, someone A meets someone B and borrows money from B. Following current development approaches, it is easy to define a $borrowMoney$ interface for B, which A then calls. However, while this design is intuitive, its underlying implication is that B is predestined to meet A and lend him money on that day; therefore, this interface has been implemented since B was born. But the real world does not work this way; the meeting of A and B is a random event determined at runtime. If circumstances change, these interfaces might either fail to connect or never be used. Change is the norm in the real world, whereas current program design solidifies these interfaces into various classes, components, and services, causing dependencies to be tightly bound and widely propagated. 

Interface-based programming inherently has these issues, as it implies a deterministic ideology\cite{LaPlace1814}. This means that during the design and development phase, it is already predetermined which modules will be called and by which modules they will be called throughout their lifecycle, with interfaces agreed upon. This does not align with the real world, just as we cannot predetermine how a newborn will interact with others throughout its life. In systems designed with this mindset, once changes occur, the world needs to be rebooted. Microservices physically isolate modules, but services still use interface-based programming, so dependency chains are also widespread.

This research adopts a different structure: substances exist independently without relying on external entities, with dependencies on the external world only embodied as conceptions originating from themselves. These conceptions of the external world are only realized at runtime, during which substances adjust and adapt to the external environment. This worldview is SCM.

The SCM adopts Aristotle's view of the world regarding substance-accident\cite{Aristotle2015organon}. Fig.~\ref{fig:substance-connections} shows the structure of this model. It holds the following views:
\begin{figure}[htbp]
  \centering
  \includegraphics[width=0.9\linewidth]{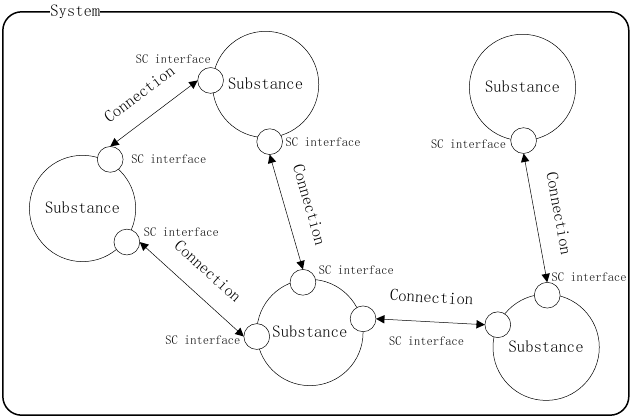}
  \caption{Substance-Connection Model}
  \label{fig:substance-connections}  
\end{figure}

\begin{enumerate}[leftmargin=*,label=\arabic*)]
\item {\texttt{Substance}}: That which endures in the essence of a thing, the fundamental unit of existence. The existence and changes of a substance do not depend on other substances.
\item {\texttt{Connection}}: The process by which substances connect with each other and form relationships. Connections depend on the substances they connect to for their existence, and the existence and changes of any substance will affect their existence and changes. The emergence of connections allows substances to acquire meaning with respect to each other. Connections coordinate the ways in which substances interact with each other.
\item {\texttt{Accidentality}}: Substances are self-existent, while connections are formed by accidental combinations of substances and change due to changes of these accidental combinations. Connections are unstable.
\item {\texttt{System}}: A system is a whole formed by a set of substances combined through connections. Because connections are accidental and variable, substances and their combinations can also change, so systems are also unstable and constantly evolving. System changes are reflected in the changes of the substances and connections that constitute it.
\end{enumerate}

The substance corresponds to the module, and the system corresponds to the application. According to the above definition of substance, any substance is absolutely independent of the system, so SCM is an ideal system.

\noindent\textbf{Proof}. Since modules interact with each other only through connections, and connections can run arbitrary functions. Assume there are two arbitrary modules $M_i$ and $M_j$, and their connection $L_{i,j}$, $M_i$ sends a request to connection $L_{i,j}$ and expects a response, while $M_j$ expects a request and can provide a response. Then, $L_{i,j}$ can run a function that receives $M_i$'s request, converts it into a request expected by $M_j$, receives $M_j$'s response, and converts it into a response expected by $M_i$. Then for any changes in $M_i$ and $M_j$, they will not impact each other. Therefore, $M_i$ and $M_j$ are completely independent. Similarly, it can be proved that all modules are completely independent of each other, i.e., each module is absolutely independent. Therefore, this type of system is an ideal system. The proof is complete. $\square$

The implication of the above proof is that $S_{mid.self}$ and all $S_{mid.interface}$.interface are invisible to the outside world, no $R(x)$ is triggered by them, and their changes will not impact other modules.

Indeed, SCM emphasizes interaction rather than manipulation, accidental connections rather than necessary dependencies, realtime coordination rather than advance prediction, inherently implying the need to reconcile substances at runtime.

The next issue to consider is how these independently developed substances can be assembled together to form a system. For the SCM architecture, this means how to minimize the cost of the functions running on the connections.

\subsection{Similar-Concept Interfaces (SCI)}

If the modules $M_i$ and $M_j$ at both ends of connection $L_{i,j}$, $M_i$'s request to connection $L_{i,j}$ is exactly what $M_j$ expects, and $M_j$'s response is exactly what $M_i$ needs, the cost of the connection is zero. This reconciliation is idealized. However, if $M_i$'s request is irrelevant for $M_j$, $L_{i,j}$ will have to run a function that is separate from $M_j$ and completely meets the needs of $M_i$. This function is named a \textbf{pseudo-module}. This is the worst case, with a cost equivalent to creating a new module while retaining the original one. In the course of this research, costs are mostly between these two extremes.

\begin{figure}[htbp]
  \centering
  \includegraphics[width=0.8\linewidth]{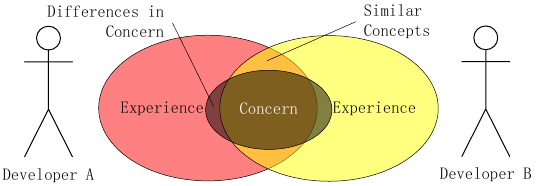}
  \caption{Similar Concepts in Concern}
  \label{fig:differences-in-concern}  
\end{figure}

How to minimize the reconciliation cost of connections is one of the core issues of this type of system. There is currently no good metric currently available for this issue. However, this research speculates that the key lies in how developers can generate similar thinking during independent development. As shown in Fig.~\ref{fig:differences-in-concern}.

Wittgenstein once raised the issue of family resemblance in his philosophical investigations\cite{Wittgenstein1953WITPI}, suggesting that words do not possess an essentially fixed meaning but rather manifest as mutual similarities, with the degree of resemblance varying from person to person. In fact, the reason for this phenomenon lies in the overlap and differences in the experiences of different individuals. Words are merely names given to the concepts formed from people's experiences, since their experiences differ, the meanings of their words naturally differ as well. From another perspective, this also implies that individuals with a large number of similar experiences will naturally develop highly similar concepts and thoughts when using the same words. This opens up the possibility for developers, who are isolated from each other, to produce interoperable modules. This is also another philosophical cornerstone of this research.

There are similar experiences between different developers. These shared experiences (orange part) are the basis for their collaboration. When they consider the same area of concern, their Similar-Concepts (SC) are the intersection of these experiences on that area of concern (orange-gray part). However, due to different experiences, these intersections also have differences. Communication can be used to reconcile these differences.

Nevertheless, areas of concern are real, while experience is inherent in developers. Alternatively, they can directly express their concern based on their experience, leaving the reconciliation process to the runtime connection. This allows developers to achieve decoupling, and modules become more independent. This is the fundamental principle of SCM and the reason why reconciliation exists.

Differences often arise not entirely because developers lack shared experience, but because they choose different concepts to describe concerns in shared experience (that is, they choose different points in the orange part). If the scope of concept selection can be limited, it is beneficial to generate SC and thus reduce the connection cost. For example, for the function of obtaining all files under a certain directory, one developer thinks that a list should be returned, and each item represents a file, including the file name and content; while another developer thinks that a map should be returned, and each key (file name) in the map corresponds to a value (content). This will result in reconciliation costs. If it is stipulated that only map or list can be used to describe the functionality, then they are more likely to generate SC.

As a result, this research introduces a development pattern based on a fixed set of universal interfaces, referred to as Similar-Concept Interfaces (SCI). It is based on the epistemology of empiricism\cite{russell1984history}, with the following propositions:
\begin{enumerate}[leftmargin=*,label=\arabic*)]
\item {\texttt{Independence}}: An individual's cognition of a substance is independent of each other, and its logical expression (language or code) does not need to be influenced by the outside world. Individuals have the ability to express a substance separately, independently of specific systems and connections.
\item {\texttt{Similar-Concepts}}: Humans are aggregates of experience. An individual's cognition of a substance is shaped by the sum of past relevant experiences. Following the principles of empiricism, individuals with the same or similar relevant experiences will have the same or similar cognitions of a substance. The higher the similarity of experience, the higher the similarity of cognition generated. This similarity of cognition is named Similar-Concepts (SC).
\item {\texttt{Connection}}: Connections between substances exist after the substances themselves. Substances with SC can establish connections. Otherwise, they cannot. The higher the degree of SC, the lower the cost of connection.
\item {\texttt{Constraint}}: When expressing a substance interacting with the outside world, completely free expression methods will affect the formation of SC. To reduce reconciliation costs, it is necessary to provide limited expression methods (interfaces) to constrain expressions. Restricted expression methods can translate SC into similar expressions (language or code).
\end{enumerate}

This research has designed a set of universal SCI. These interfaces divide all substances into two categories, resource and process, when describing the relationship between substances. This approach considers all computation and processing of the system as the result of processes running on resources. At the same time, the following conventions are made for the development process of modules in terms of methodology:
\begin{enumerate}[leftmargin=*,label=\arabic*)]
\item {\texttt{Isolation}}: After separating the concerns in the system, modularization should be carried out. Developers should focus on the concern itself rather than negotiating interfaces with other developers. The development process should be as independent as possible, and communication between developers should be minimized.
\item {\texttt{Concern-centric}}: Developers describe based on their own cognition of the concern, considering as much as possible the general SC on the concern.
\item {\texttt{Connection description}}: For descriptions not included in the current concern (including but not limited to objects, methods, or processes, which are usually connected to other modules), developers should use one or more interfaces defined by SCI to refer to them. For any module, if it provides services to the outside world, it should implement one or more interfaces defined by SCI for external use.
\item {\texttt{Input and output}}: Except for basic data structures (generally provided by the base library of the development language), any custom data structures that need to be input or output should implement one or more interfaces defined by SCI for external use.
\end{enumerate}

It can be seen that module division of SCM is based on such universal interfaces, which is also in stark contrast to Interfaces-based Module Division. Therefore, SCM can also be called Universal Interfaces-based Module Division. Based on the above concepts, this research developed a new architecture.

\section{Instantiation}

When it comes to the technical implementation of the ideal system, it is necessary to consider the aforementioned module impact scope within three types of contexts. To be absolutely independent in the static context, it is necessary to use Universal Interfaces-based programming for module division and development; whereas to be absolutely independent in the runtime context, the system must provide a dynamic runtime execution environment for the independent loading, execution, and updating of each module; and to be absolutely independent in the non-runtime context, it is also required that the software development process and team structure revolve around modules, while minimizing connections between teams.

The SCM can be implemented using Kubernetes based Service Mesh\cite{ServiceMesh2024}, such as Envoy\cite{Envoy2024}, which is a good location of pseudo-modules. However, it is not necessary to use such a heavy framework. This research fully implemented it using Java, and this platform is named EIGHT, which uses OSGi\cite{osgi2024} and Spring as the infrastructure. 

Merely changing the technical architecture is not sufficient to ensure module independence. According to Corollary 2, coupling between modules is caused by developers' mutual understanding. This research adopted new project management methods in subsequent experiments: splitting a team of about 8 developers into 3-4 groups, assigning development tasks separately, isolating their communication, and allowing each group to independently decide on module design and development, without constraining the development order. A review meeting is convened only when interface mismatches occur during assemble time to determine the coordination approach.

\subsection{The Architecture of EIGHT}
\begin{figure}[htbp]
  \centering
  \includegraphics[width=\linewidth]{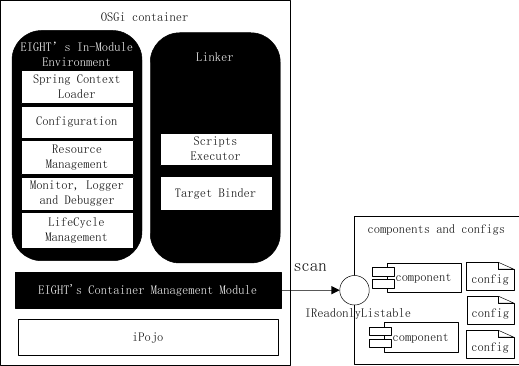}
  \caption{The Main Parts of EIGHT}
  \label{fig:main-parts}  
\end{figure}

EIGHT chose the Felix\cite{felix2024} container and used the iPojo composition layer\cite{ipojo2024}. The traditional OSGi framework has so many explicit or implicit dependencies that it is difficult to construct an ideal system that allows modules to be changed arbitrarily. However, it is different after implementing SCM. EIGHT designed the architecture based on SCM and SCI and implemented it using OSGi's module dynamic loading technology. It consists of a set of SCIs (15 interfaces and 14 methods, as shown in Tab. ~\ref{tab:SC-interfaces}), a set of base libraries, and several runtime modules (as shown in Fig.~\ref{fig:main-parts}).
\begin{table}[htbp]
	\renewcommand{\tablename}{Tab.}
	\caption{The Similar-Concept Interfaces}
	\begin{center}
		\begin{tabular}{|l|l|}
		\hline
		Interface&Description\\
		\hline
    		IProcessor &Unary process function\\
		\hline    		
    		IBiProcessor&Binary process function\\
		\hline    		
    		ITriProcessor&Trinary process function\\
    		\hline
    		IInputResource&Read-only resource\\
    		\hline
    		IOutputResource&Write-only resource\\
    		\hline
    		IListable&Listable resource\\
    		\hline
    		IResource&Read-write resource\\
    		\hline
    		IReadonlyListable&Listable read-only resource\\
    		\hline
    		IListableResource&Listable read-write resource\\
    		\hline
    		ITransaction&Resource supporting transactions\\
    		\hline
    		ITransactionResource&\makecell[l]{Read-write resource supporting\\transactions}\\    
    		\hline
    		IListableTransaction&\makecell[l]{Listable read-write resource\\supporting transactions}\\    
    		\hline
    		IExtendable&Extension interface\\
    		\hline
    		IThing&Thing interface\\
    		\hline
    		IUniversal&The sum of all interfaces\\ 
		\hline
		\end{tabular}
		\label{tab:SC-interfaces}
		\end{center}
\end{table}

The runtime modules have the following functions:
\begin{enumerate}[leftmargin=*,label=\arabic*)]
\item {\texttt{Container Management Module}}: A set of services loaded when the OSGi container starts. The main functions include providing container-level shared resources, monitoring and managing running instances, scanning and loading components and configurations, etc. 

\item {\texttt{In-Module Environment}}: A runtime environment within the module provided by the base libraries. Each instance created by component needs to own one and configure it during instantiation. It is mainly responsible for managing the module's lifecycle, starting and recycling the Spring Context, providing parameter configuration for the running Context, providing module-level resource accessment, access control, monitoring, debugging, logging, etc.

\item {\texttt{Linker}}: In EIGHT, the linker is a special component that can generate different instances with different configurations, used to bind targets and connect instances. The linker is responsible for attaching the connection points to the specified services of another instance according to the configuration. The linker also implements the functions in SCM and can be configured with scripts (currently supporting groovy\cite{groovy2024}), classes or packages that implement specific interfaces. These scripts, classes or packages will be executed when it is invoked.
\end{enumerate}

\begin{figure}[htbp]
  \centering
  \includegraphics[width=\linewidth]{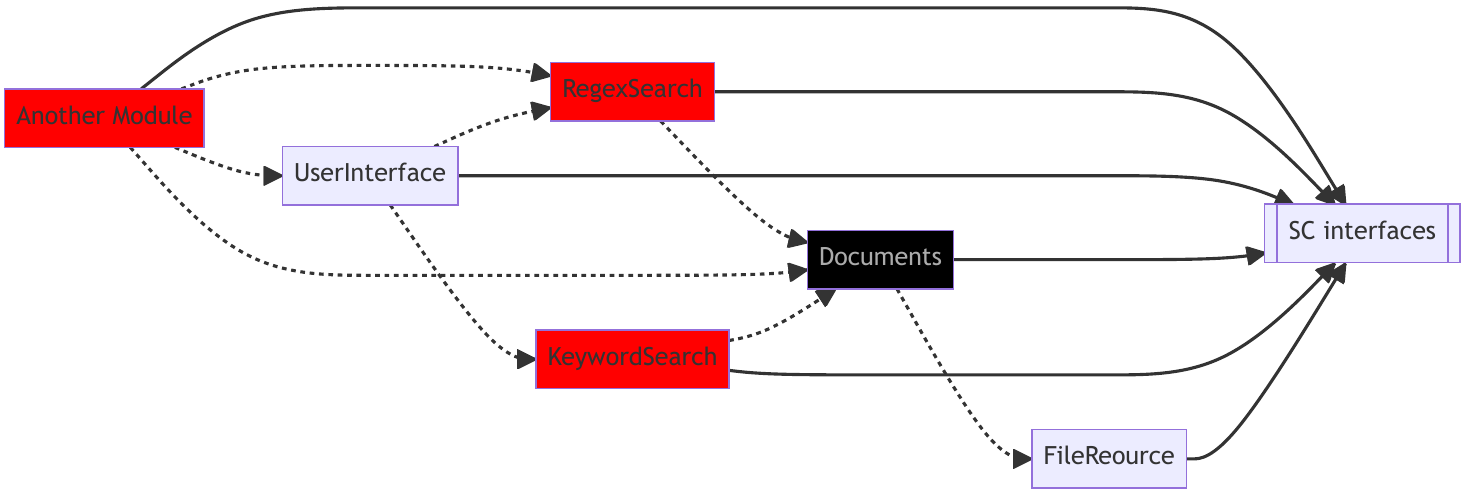}
  \caption{Depends on Similar-Concept Interfaces}
  \label{fig:depend-on-sc-interfaces}  
\end{figure}

SCI not only plays a role in constraining developers' similar-concepts in EIGHT but also serves as a decoupling function for static dependencies from a technical perspective, turning mesh dependencies into tree dependencies, with the root being the constant SCI. The development model oriented to SCI makes the components naturally independent, making it easy to unload and change in the OSGi container. As shown in Fig.~\ref{fig:depend-on-sc-interfaces}, the dashed lines represent the dependency relationships based on the interfaces provided by each module under the traditional OSGi, with the black modules being the changing modules and the red modules being the affected modules. After switching to SCI, there are no explicit dependencies between modules, so changes in the black modules will not cause the red modules to reload. Similarly, it also removes the static dependencies of pseudo-modules on both ends of the connection, greatly helping to release the runtime resources.

EIGHT is a never-ending process at runtime, mainly performing the following operations:
\begin{enumerate}[leftmargin=*,label=\arabic*)]
\item {\texttt{Loading, updating, and unloading}}: The EIGHT platform periodically scans for changes in components and configurations, compares them with those in the container, and then loads, updates, and unloads components and configurations.
\item {\texttt{Instance configuration}}: The EIGHT platform generates instances of components based on configurations and injects configurations into the instances.
\item {\texttt{Linkers}}: The EIGHT platform generates linkers based on configurations, configures them, and then connects the instances into a system.
\item {\texttt{Monitoring and management}}: The EIGHT platform provides monitoring and management functions, observing the status of each instance at runtime and manually controlling and modifying configuration parameters for instances.
\end{enumerate}

\subsection{A Simple Example of EIGHT}
Returning to the previous search application, the core code of the UserInterface and Search modules is as follows:
\begin{lstlisting}
public class UserInterface implements IProcessor<String, String> {
	protected IProcessor<String, Integer> finder;
	protected IProcessor<String, String> formatter;	
	
	public String process(String keyword){
		return formatter.process(
			finder.process(keyword).toString());
	}
}

public class Search implements IProcessor<String, Integer> {
	protected IListable<Object, String> documents;

	public Integer process(String keyword) {
		Map<Object[], String> docs = documents.all();
		//Iterate through the file to count keywords
		return count;
	}
}
\end{lstlisting}

Coming back to that previous issue. If the functionality of the Search module changes and an additional directory parameter needs to be provided, for example:
\begin{lstlisting}
public class Search implements IBiProcessor<String,String, Integer> {
	protected IListable<Object, String> documents;

	public Integer perform(String keyword, String dir) {
		Map<Object[], String> docs = documents.all(dir); //Retrieve all files in the dir.
		//Iterate through the file to count keywords
		return count;
	}
}
\end{lstlisting}

For EIGHT, at this point, it is only necessary to adapt with a script on the linker, where the property $context$ can be used to get the connection points on both ends of the linker.
\begin{lstlisting}
// userinterface-search.groovy
class Proxy implements IProcessor {
    IProcessor context
	
	def process(Object keyword) { 
		def dir = "searchPath"
		def ret = ((IBiProcessor)context.process("next"))
						.perfrom(keyword, dir)
		ret
	}
}
\end{lstlisting}

This also aligns more with the logic of the real world: in principle, the Search module cannot foresee that the connected module is searching files from a directory and has no reason to provide a directory parameter. If the Documents module makes an unreasonable additional request, it can only negotiate a $searchPath$ when the connection occurs, and the Search module is not obliged to solidify it in its own logic. This simple script demonstrates a basic way to bridge Similar Concepts.

\begin{figure}[htbp]
  \centering
  \includegraphics[width=0.9\linewidth]{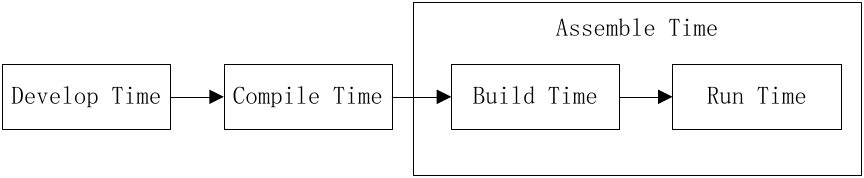}
  \caption{Assemble Time}
  \label{fig:assemble-time}  
\end{figure}

In fact, the development approach of EIGHT differs from both the waterfall model and agile development. It views software as a continuously changing process, with these changes occurring through connections between different entities. Corresponding to the connection process, there is an important tense in the software lifecycle: assemble time. This tense generally occurs after compilation, during system construction, and covers the entire process of operation, maintenance, and iteration, including runtime. As shown in Fig.~\ref{fig:assemble-time}.

\subsection{A Formal Example of EIGHT}

In addition to the above conceptual example, EIGHT has also been used to develop several formal enterprise applications. One such case employs the Spring MVC framework. Unlike the standard Spring framework, it cuts all business lines vertically and then cuts Model, View and Controller horizontally. This forms a basic module grid, with additional connection modules providing links between businesses. The front-end also provides a unified and extensible UI framework.

\begin{figure*}[htbp]
  \centering
  \includegraphics[width=0.7\linewidth]{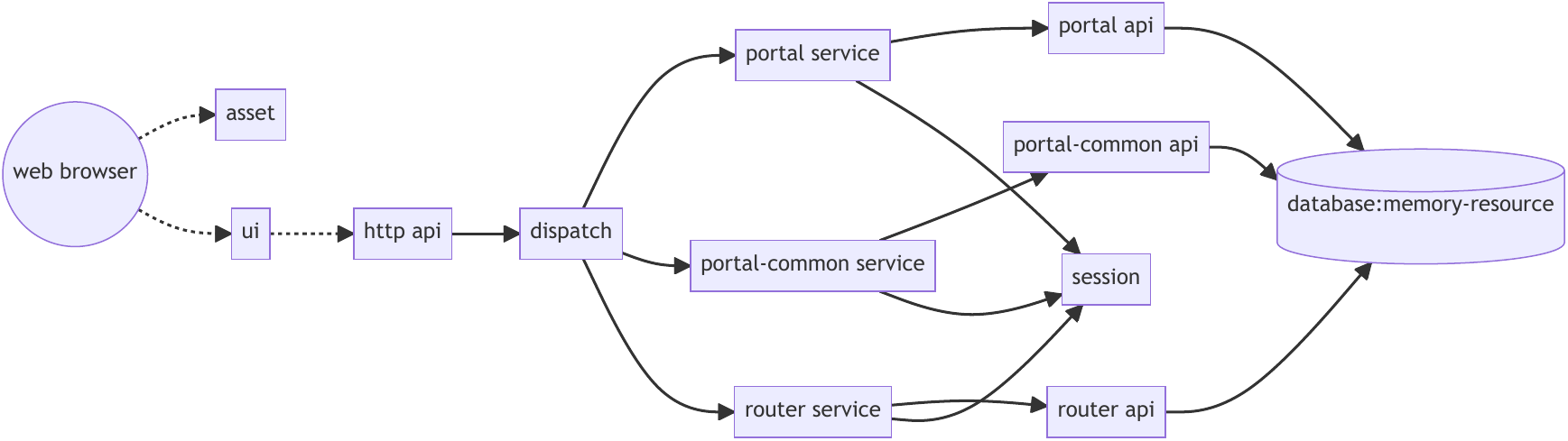}
  \caption{The Module Structure of the Router Management Demo}
  \label{fig:another-demo-structure}  
\end{figure*}

As shown in Fig.~\ref{fig:another-demo-structure}, this is a framework with basic user management and authentication, on which any business line can be added, providing various foundational services to these businesses. In the current structure, a Router Management business unit is provided.

\begin{figure*}[htbp]
  \centering
  \includegraphics[width=0.8\linewidth]{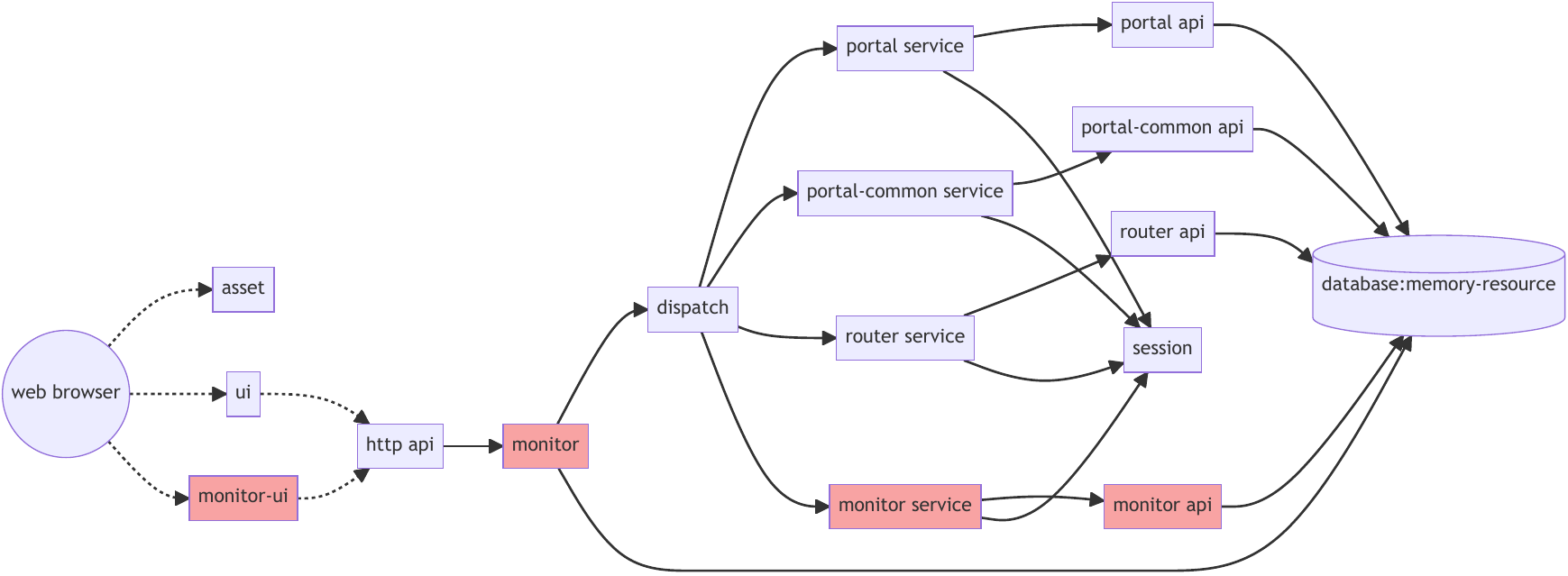}
  \caption{The Module Structure of the Router Management Demo with Monitor Modules}
  \label{fig:another-sample-with-addtional-module}  
\end{figure*}

Within the entire framework, any module is able to change dynamically at runtime. Therefore, these business lines can change the business logic at any time as needed, or even dynamically load or unload entire business lines or the entire application. As shown in Fig.~\ref{fig:another-sample-with-addtional-module}, a new business is embedded into the existing system to perform interception and monitoring functions. Such functions, which require extensive adjustments to service routing and network topology even with microservices, can be easily loaded at runtime under the EIGHT framework without any impact on the ongoing business operations.

In terms of software development organization, these modules are divided among various groups according to business lines, with each group developing in parallel while maintaining communication isolation, only discussing and reviewing the differences in their similar concepts at assemble time to determine a harmonization solution.

These reveal that module independence, parallel iteration of development, and dynamic changes in applications do not necessarily require physical isolation. By adopting a reasonable worldview and methodology to design systems and evolve architectures, even within the same process, all these can be achieved.

All above examples can be obtained from the external link\cite{SupplementaryMaterial}, which provides the complete system platform, documentation, and examples. The supplementary material also provides more examples, including several robot control applications based on ROS2\cite{ros2024}. Users can perform various operations on these applications according to the documentation, change their parts at runtime, and even change the entire system.

\subsection{The Superiority of EIGHT}
In addition to supporting better modularity and development independence, EIGHT also exhibits excellent technical characteristics. Due to the fundamental and significant differences between architectures, this research only makes limited comparisons based on above several examples to explain the main differences. This research implemented the same functions and module structures using microservices and EIGHT, as shown in Tab. ~\ref{tab:the-comparison-of-microservices-and-eight}.
\begin{table}[htbp]
	\renewcommand{\tablename}{Tab.}
	\caption{The Comparison of Microservices and EIGHT}
	\begin{center}
		\begin{tabular}{|c|c|c|}
		\hline
		&Microservices&Component Platform\\
		\hline
    		Hardware &\makecell[c]{Dell R410 \\intel E5506×2 \\128GB ram}& \makecell[c]{Qualcomm arm64\\Snapdragon 410\\512MB ram} \\
		\hline    		
    		Environment&Kubernetes 1.19&$\geq{}$JRE 1.6.14 \\
		\hline    		
    		Language&goLang 1.11.5& Java SE 6\\
    		\hline
    		Module loading&0.1-several seconds&can be ignored \\
    		\hline
    		\makecell[c]{Modules size \\(Simple Example)}&\makecell[c]{2-5MB\\avg 2.3MB}&\makecell[c]{9-200KB\\avg 54KB}\\
    		\hline
    		\makecell[c]{Memory usage \\(Simple Example)}&Avg 4GB& Avg 120MB \\
    		\hline
    		\makecell[c]{Response time \\(Simple Example)}&$\leq{500}$ms&$\leq{100}$ms\\
    		\hline    
     	\makecell[c]{Modules size \\(Router Management)}&\makecell[c]{2-7MB\\avg 3MB}&\makecell[c]{10-770KB\\avg 90KB}\\
    		\hline
    		\makecell[c]{Memory usage \\(Router Management)}&Avg 4-5GB& Avg 210MB \\
    		\hline
    		\makecell[c]{Response time \\(Router Management)}&500ms-2s&100-500ms\\
    		\hline    
     	\makecell[c]{Modules size \\(ROS examples)}&\makecell[c]{2-6MB\\avg 3.4MB}&\makecell[c]{9-230KB\\avg 55KB}\\
    		\hline
    		\makecell[c]{Memory usage \\(ROS examples)}&Avg 600MB& Avg 110MB \\
    		\hline
    		\makecell[c]{Response time \\(ROS examples)}&$\leq{300}$ms&$\leq{100}$ms\\
    		\hline     						
		\end{tabular}
		\label{tab:the-comparison-of-microservices-and-eight}
		\end{center}
\end{table}

EIGHT has the following advantages over microservices:
\begin{enumerate}[leftmargin=*,label=\arabic*)]
\item {\texttt{Better performance}}: Although the response time difference in Tab. ~\ref{tab:the-comparison-of-microservices-and-eight} is not very significant, it is because microservices are running on a single node. In a distributed scenario, there would be a significant difference due to the cost difference between in-process stack calls and network transmission and serialization.
\item {\texttt{Less resource consumption}}:  Deploying the microservices environment required a server and occupied more than 4GB of memory, mainly due to environmental consumption (The ROS environment is an exception, considering the resource constraints of the application scenario, only docker container was deployed on the appliance without the use of the K8s).  EIGHT, however, can run in any environment with JRE 1.6 or above. The test environment was an embedded CPU, occupying about 100-200MB of memory. The difference in resource consumption affects environmental adaptability.
\item {\texttt{Module size}}: goLang modules are larger after compilation due to the built-in VM, while EIGHT and Java provide a set of base libraries for modules, with module packages usually ranging from tens to hundreds of KB. The module size affects the ability to remotely deploy and update in WAN or weak network environments.
\item {\texttt{Loading time}}: Microservices involve process restart, with loading time depending on the process release and start time; EIGHT uses a daemon thread to load and unload the system, and the time to start a component can be ignored. Notably, EIGHT loading new modules does not affect ongoing business, and the old modules are released after the ongoing requests are completed. The switching process is seamless, with no loss of process state. This is crucial for real-time systems with high SLA requirements (such as in-vehicle systems).
\end{enumerate}

In addition to the advantages reflected in Tab. ~\ref{tab:the-comparison-of-microservices-and-eight}, EIGHT has the following benefits:
\begin{enumerate}[leftmargin=*,label=\arabic*)]
\item {\texttt{Easy implementation and maintenance}}: EIGHT's simple structure and low environmental and skill requirements for deployment and implementation help to reduce the hardware, software and personnel implementation and maintenance costs of systems.
\item {\texttt{Enhanced data consistency}}: The single-process structure of EIGHT ensures transaction consistency, reduces the possibility of distributed transaction data conflicts, and lowers development difficulty.
\item {\texttt{Reduced failures}}: The single-process structure of EIGHT eliminates the uncertainties of distributed services, avoiding issues such as service failures or unavailability.
\item {\texttt{Easy tracking}}:EIGHT runs on the same node, allowing for centralized viewing of runtime module structures and logs, facilitating fault tracking and analysis.
\item {\texttt{Strong environmental adaptability}}: EIGHT is easy to deploy, consumes minimal resources, and has the ability to continuously adapt. This makes it possible to apply to areas that microservices and monolithic applications struggle, providing new solutions for system integration, dynamic deployment, changes and upgrades, and remote management and maintenance in resource-constrained environment, legacy systems, edge devices, discrete nodes, etc.
\end{enumerate}

Beyond that, the adoption of new software engineering methods also demonstrates unique advantages:
\begin{enumerate}[leftmargin=*,label=\arabic*)]
\item {\texttt{More efficient development organization}}: EIGHT emphasizes developer isolation and maintaining ignorance of external modules. The work of each development group is minimally dependent on each other, and interactions with external assumptions can easily be replaced with stub modules under the EIGHT framework. Thus, the development of each development group is almost entirely parallel, and communication costs are significantly reduced compared to previous development models (whether waterfall or agile), resulting in shorter module iteration cycles.
\item {\texttt{Better module reusability}}: Since module developers focus on the functionality of the module itself and cannot foresee external interfaces and interaction methods, they tend to conceive as many usage scenarios as possible. This not only results in better module cohesion but also provides better external adaptability, making modules easier to reuse in more environments.
\item {\texttt{More flexible module granularity}}: Under the EIGHT technical architecture, the additional overhead of splitting modules is minimal, making it easier to segment software design according to its business logic boundaries without considering costs. This also enhances software engineering organizational efficiency and module reusability. 
\end{enumerate}																	

\begin{figure}[htbp]
  \centering
  \includegraphics[width=0.8\linewidth]{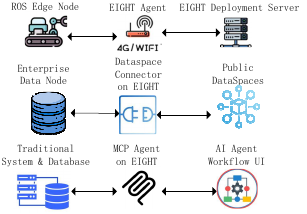}
  \caption{Engineering Practices of EIGHT}
  \label{fig:eight-engineering-practices}  
\end{figure}

Currently, EIGHT has been applied in several engineering fields and has achieved excellent practical results. As shown in Fig.~\ref{fig:eight-engineering-practices}.

For example, it can be deployed flexibly across various environments, offering a suitable solution for edge device deployment and remote updates in environments with weak network connectivity, such as wide-area systems using the mobile internet;
It offers deployment and maintenance similar to monolithic applications while providing development methods akin to microservices, making it suitable for connecting dispersed internal data sources to public dataspaces;
It can be deployed cheaply and conveniently across various parts of an enterprise and can change as needed, making it suitable for integrating traditional systems with AI agents and for working as MCP services. In the future, AI systems can intelligently assemble EIGHT systems based on demand and deploy them to appropriate agent nodes to perform various tasks.

\section{Related Work}
This SCM-based system architecture and SCI-based design theory are introduced for the first time in this paper, and no related research have yet been conducted. The Istio architecture\cite{istio2024} offers a possibility for implementing SCM on microservices, but it is mainly used as a trusted proxy between services yet.

Currently, microservices are the dominant and increasingly prevalent application layer architecture, with the main trend being the migration of traditional monolithic applications to microservices\cite{Knoche2019DriversAB,Zhang2019MicroserviceAI,MichaelAyas2023AnES,Mazlami2017ExtractionOM,Taibi2019FromMS}. However, some researches focus on the negative impacts of microservices, such as complex transaction coordination\cite{Laigner2021DataMI,Ghofrani2018ChallengesOM}, performance impacts\cite{Jamshidi2018MicroservicesTJ}, consistency\cite{Furda2018MigratingEL}, multi-database manage-ment\cite{Kalske2017ChallengesWM}, upgrade failures\cite{Zhang2021UnderstandingAD}, automation in distributed environ-ments\cite{10.1007/978-3-319-64218-5_40} and difficulty in understanding the system as a whole\cite{Simhandl2023DevelopersCE}. Others focus on the implementation surveys of microservices, aiming to provide evaluable recommendations for their applicability\cite{Zhou2022RevisitingTP,Taibi2017ProcessesMA,Velepucha2023ASO}.

Some improvements to microservices are also starting to emerge. For example, Ghemawat\cite{Ghemawat2023TowardsMD} suggests that microservices should separate logical services from physical deployments. Whereas, Some other researches focus on the refactoring of microservices, automated generation of services, and so on \cite{Waseem2022DecisionMF,Perera2018ARS,Anand2023BlueprintAT}.

OSGi had a number of applications in areas such as server-side and mobile Internet before microservices existed\cite{Zhang2013TowardsAO,Verbelen2011DynamicDA,Rellermeyer2007ROSGiDA,Zhang2014AnOF}.Nowadays, OSGi is less researched and mainly focused on areas such as IoT\cite{Bixio2020AFI,Mamro2020MICROSERVICESB}.

The above studies have not explored the principles of modularity and the direction of architectural evolution in terms of programming thinking and software engineering methodologies, which significantly differ from this paper.

\section{Conclusion}
In this paper, we make the following contributions:
1) A set of metrics was designed to measure the degree of inter-module coupling and its impact scope.
2) To further improve independence, a conceptual architectural model is proposed.
3) Describe the fundamental principles and development methodology for the conceptual architectural model described above.
4) An implementation of the architecture is presented and examples are provided. 

\bibliographystyle{ACM-Reference-Format}
\bibliography{eight}
\end{document}